\documentclass[prl,twocolumn]{revtex4}
\usepackage{graphicx}% Include figure files
\usepackage{dcolumn}% Align table columns on decimal point
\usepackage{bm}% bold math

\begin{document}

\preprint{Preprint}
\bibliographystyle{prsty}

\title{Random Networks Growing Under a Diameter Constraint}
\author{ Rajan M. Lukose and Lada A. Adamic}
\address{Information Dynamics Lab, Hewlett-Packard Laboratories, 1501 Page Mill Road, CA 94304-1126}

\begin{abstract}
We study the growth of random networks under a constraint that the
diameter, defined as the average shortest path length between all
nodes, remains approximately constant. We show that if the graph
maintains the form of its degree distribution then that
distribution must be approximately scale-free with an exponent
between $2$ and $3$. The diameter constraint can be interpreted as
an environmental selection pressure that may help explain the
scale-free nature of graphs for which data is available at
different times in their growth.  Two examples include graphs
representing evolved biological pathways in cells and the topology
of the Internet backbone.  Our assumptions and explanation are
found to be consistent with these data.
\end{abstract}

\pacs{Valid PACS appear here}% PACS, the Physics and Astronomy
\maketitle

Measurements on a wide variety of networks such as the World Wide
Web\cite{diameter,ladamicsw}, the Internet backbone
\cite{achilles_heel,faloutsos99topology}, social networks
\cite{chung,strogatz_review,wattssw}, and metabolic
networks\cite{barabasi,fell} have shown that they differ
significantly from the classic Erdos-Renyi model of random graphs
\cite{erdos}. While the traditional Erdos-Renyi model has a
Poisson node degree distribution, with most nodes having a
characteristic number of links, these networks have highly skewed,
scale-free degree distributions approximately following a power
law $p(k) \sim k^{-\gamma}$, where $k$ is the node degree, and
$\gamma$ is the scale-free exponent. To account for these
observations, random graph growth models have been developed
\cite{huberman_growth,pref_attatch,stoch_web,krapivsky2000,dorogovtsevPRE2000}
that rely on the intuitively appealing idea of preferential
attachment.

For the most part, these models are of growing networks in which
new nodes are added to a graph by addition of one or more edges to
already existing nodes. The attachment is preferential because the
likelihood of attachment to a node depends on the number of other
nodes already linked to it.  Thus, such models rely solely on
endogenous factors, since they do not take account of any global
exogenous selection pressures which might shape the form of
evolving and growing networks.

Such selection pressures would be especially relevant, for
example, in a biological context. Measurements of the topological
properties of graphs representing the metabolic networks of $43$
organisms have demonstrated their scale-free nature
\cite{barabasi,metabolicproceedings}.  In all of these graphs, of
different sizes, it was found that the diameter, defined as the
average shortest path between every pair of nodes in the graph,
was constant.  Such a constancy may be related to important
properties relevant to the core functioning of these biological
networks, such as the spread and speed of responses to
perturbations \cite{fell,fellproceedings}.

Another example is the Internet backbone graph, whose growth and
evolution over time has been studied by several authors
\cite{faloutsos99topology,vespignaniPRLbackbone,goh2002PRLtopology}.
In this case, there are performance and robustness constraints
that such networks much satisfy. These constraints can be thought
of as environmental pressures (which may operate indirectly) that
would select against highly inefficient network structures.  One
possibility may be a bias in favor of network changes and
additions that tend to maintain the average shortest path.

These two cases are relatively rare examples of network data
representing graphs of varying size shaped by similar selection
pressures, and they allow the testing of explanations for their
generic features. The main selection-based explanation
\cite{barabasi} for scale-free network topologies relies on the
fact that such networks are robust with respect to random
malfunction of nodes \cite{achilles_heel}. Robustness is
identified with the diameter of the network, and scale-free
networks maintain their diameter when nodes are eliminated at
random.  However, while scale-free graphs are robust in this
sense, it has not been shown that robust graphs must necessarily
be scale-free.

In this Letter, we argue that another notion of robustness can be
added to the error tolerance argument. Namely, scale-free networks
are special in the sense that they can grow, with the same
functional form for the degree distribution, and simultaneously
maintain an approximately constant diameter. This implies that
when graphs grow and evolve in an environment in which there are
selection pressures on the diameter, these graphs are likely to be
scale-free.  Our results help clarify the connection between the
apparent constancy of the diameter and the scale-free topology of
the graphs in the two examples studied.

We consider random graphs of different sizes under the constraint
that the diameter remain approximately constant as the graph
grows. We show that if the graph maintains the form of its degree
distribution then that distribution must be approximately
scale-free with an exponent between $2$ and $3$. These results may
help explain the scale-free nature of graphs, of varying sizes,
representing the evolved metabolic pathways of different organisms
and the topology of the Internet backbone. Our assumptions and
results are consistent with empirical findings.

We start with an expression for the diameter $d$ of a random graph
with arbitrary degree distribution developed in Ref.
\cite{newman01graphs} using a generating function formalism
applied to the degree distribution $p\left(k\right)$,

\begin{equation}\label{newman}
  d = \frac{ \log N }{ \log(z_{2} / z_{1}) } -   \frac{ \log z_{1} }{ \log(z_{2} / z_{1}) } + 1
\end{equation}
This formula depends only on the number of nodes in the graph $N$,
the average degree of the nodes $z_{1}$, and the average number of
nodes $z_{2}$ that are reachable in two edge traversals from an
arbitrary node.  The formula was derived by considering an
ensemble of random graphs (without explicit clustering, see Refs.
\cite{newman01graphs,newmanchapter} for details), and calculating
from the degree distribution the average number of nodes within
some radius of edges away from a random node.  As the radius gets
larger, the number of nodes enclosed grows rapidly. When that
number of nodes is approximately equal to $N$, the corresponding
radius is a good approximation to the average shortest path in the
graph, since most of the nodes are at that distance. If the random
graph is directed, the same argument applies, and Eq.
(\ref{newman}) is valid when $p\left(k\right)$ is the degree
distribution of outgoing edges.

It is important to emphasize that Eq. (\ref{newman}) is an
approximate formula for the average shortest path of a random
graph (without clustering or assortative mixing, etc.).  While the
formula is not numerically precise, it does capture the dependence
of the diameter on $N$\footnote{For example, a recent rigorous
analysis for a slightly different random graph model in
\cite{chungPNASdiameter} arrived at a very similar formula to the
one we use, but with a leading constant factor. In
\cite{newman02social}, the reasonable accuracy of Eq. \ref{newman}
is demonstrated for various real-world networks.}.

To calculate the diameter of a graph with degree distribution
$p\left(k\right)$, but with a finite number of nodes $N$, the size
of the graph must be parameterized through the degree
distribution. In any fully connected finite-sized graph, the
smallest possible degree is $1$ and the largest degree in the
graph must be less than $N$. The parameterization can be
accomplished by imposing an $N$-dependant cutoff function
$k_{c}\left(N\right) < N$ and writing $\left< k^{i} \right> =
\sum_{1}^{k_{c}\left(N\right)} k^{i} p\left(k\right)$.

Additionally, a simple application of the generating function
method leads to the relationship $z_{2}= \left< k^{2} \right> -
\left<k\right>$ ($\left<k\right> = z_{1}$ by definition), which
allows the diameter to be calculated from the first two moments of
$p\left(k\right)$.

We seek a degree distribution that maintains its functional form
and has an approximately constant diameter independent of $N$.
Effectively, this means a single function $p\left(k\right)$ that
does not depend on $k_{c}$ except through its normalization. (We
rule out the most obvious example, a star-shaped graph with $N-1$
nodes each connected to the $N$th node.  This graph, whose
construction is deterministic, has an approximately constant
diameter of $2$.)

Returning to Eq. (\ref{newman}), consider the first and second
terms.  The first plus the third term are an upper bound on $d$
since the second term is strictly non-negative and always $\leq
1$. In addition, for any fixed $p\left(k\right)$, the second term
is non-increasing as $k_{c}$ gets larger, and will approach a
constant $\leq 1$ if $p\left(k\right)$ has finite moments as $N
\rightarrow \infty$. As a result, for our purposes the first term
is dominant, and we will neglect the second.  If the first term is
a constant $c$, then the diameter will always lie between $c$ and
$c+1$.

Thus, in order to find the degree distribution with approximately
constant diameter independent of $N$, we set the first term in Eq.
(\ref{newman}) be equal to a constant $c$, which results in the
requirement that

\begin{equation}\label{constdiam}
N^{1/c} = z_{2}/z_{1}
\end{equation}
The ratio $z_{2}/z_{1}$ is also the average degree of a node found
by following a randomly chosen edge \cite{newman01graphs}.  This
average degree must always be less than the largest degree in the
graph, which provides a lower bound on the cutoff function
$k_{c}\left(N\right)$, resulting in $N^{1/c} < k_{c}\left(N\right)
< N$ for all $N$. The function $k_{c}\left(N\right)$ is therefore
bounded from above and below by power functions and we will use
the explicit form $k_{c}\left(N\right) = N^{\alpha}$ where $1/c <
\alpha \leq 1$.  Setting $\alpha = 1$, we recover the least
restrictive case where no node has a degree greater the size of
the graph.  Letting $\alpha$ vary also allows us to see the
consequences of different cutoff dependencies if they are
consistent with empirical findings.

\begin{figure}
\begin{center}
\includegraphics[scale=0.4]{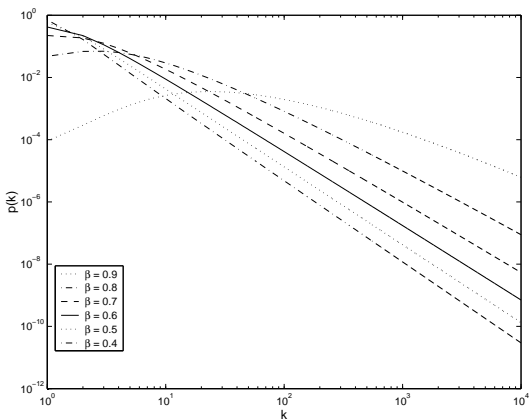}
\end{center}
\caption{A plot of the numerically calculated distributions for
various values of $\beta$ and with a
$k_{max}=10^4$.\label{theodists}}

\end{figure}

Then, using again the assumption $N \gg 1$, Eq. (\ref{constdiam})
can be written as an equation for $p\left(k\right)$ involving the
ratio of its moments:

\begin{equation}\label{constdiamsum}
k_{c}^{\frac{1}{\alpha c}} = \frac{\sum_{1}^{k_{c}} k^{2}
p\left(k\right)}{\sum_{1}^{k_{c}} k p\left(k\right)}
\end{equation}
The distribution $p(k)$ can be determined by writing this equation
for $k_c$ and $k_c\pm 1$.  Algebraic manipulation yields the
recurrence relation

\begin{equation}\label{recurrence}
    p(k+1) = p(k) \frac{k^2 - k (k-1)^\beta}{(k+1)^2 -
    (k+1)^{\beta+1}} \frac{(k+1)^\beta - k^\beta}{k^\beta - (k-1)^\beta}
\end{equation}
where $\beta \equiv 1/\alpha c$ for convenience.  The full
solution can easily be calculated numerically.  For most values of
$\beta$ the distribution approaches its scale-free asymptote
rather quickly (by $~k=100$), as shown in Fig. \ref{theodists}.

A more explicit analytic characterization of the solution can be
found by using an integral approximation for Eq.
(\ref{recurrence}), $k_{c}^{\frac{1}{\alpha c}} =
\frac{\int_{1}^{k_{c}} k^{2} p\left(k\right) dk}{\int_{1}^{k_{c}}
k p\left(k\right) dk}$ This integral equation can be solved
exactly by turning it into a first-order differential equation for
the function $p\left(\cdot\right)$ by differentiations with
respect to $k_{c}$. Replacing $k_{c}$ with a dummy variable $k$,
and writing $\beta = 1/\alpha c$ for convenience, the resulting
equation is $((\beta - 3) k + (\beta + 2)
k^{\beta})p\left(k\right) + k (k^{\beta} - k) p'\left(k\right) =
0$. The unique solution, up to a normalization constant dependent
on $k_{c}$, is
\begin{equation}\label{solution}
p\left(k\right) = \frac{1}{k^{3 - \beta}} (1 - \frac{1}{k^{1 -
\beta}})^{\frac{1 - 2 \beta}{\beta - 1}}
\end{equation}
The solution is in good agreement with the result of numerical
iteration of the exact result of Eq. (\ref{recurrence}) away from
$k=1$. The singular behavior for $\beta < 1/2$ at $k = 1$ is the
result of the continuous approximation.  Nevertheless, Eq.
(\ref{solution}) well approximates the form of the distribution
with an approximately constant diameter that is always between $c$
and $c + 1$, approaching $c + 1$ asymptotically with $N$. For $k
\gg 1$, it is also a scale-free distribution with exponent $\gamma
= 3 - 1/\alpha c$. Since $\alpha > 1/c$ and $c \geq 1$ is a
natural bound on $c$, the scale-free exponent satisfies $2 <
\gamma < 3$, and it can be written in terms of the diameter $d$
and the cutoff exponent $\alpha$ as $\gamma = 3 - \frac{1}{\alpha
(d - 1)}$.

\begin{figure}
\begin{center}
\includegraphics[scale=0.5]{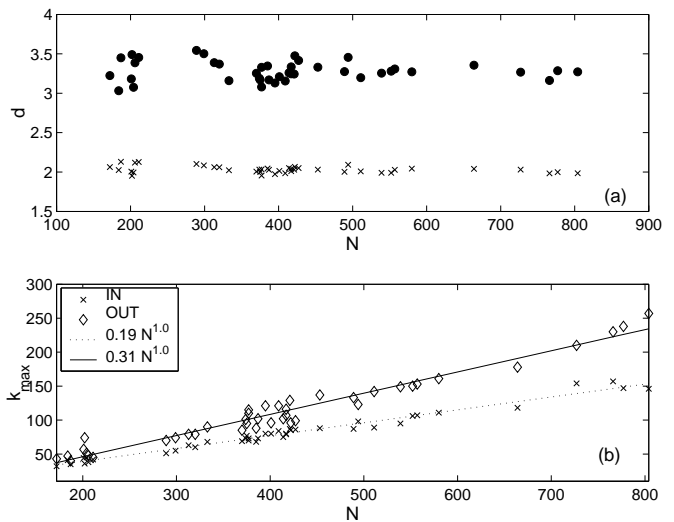}
\end{center}
\caption{(a) The diameter $d$ is almost constant as a function of
$N$, the size of the metabolic network. The top set of points are
the measured average shortest path on the graphs, while the bottom
set shows the quantity computed from the degree distributions
using Eq. (\ref{newman}). (b) The maximum in degree and out degree
of the node scales linearly with the size of the network.}
\label{metabolic}
\end{figure}

\begin{figure}
\begin{center}
\includegraphics[scale=0.5]{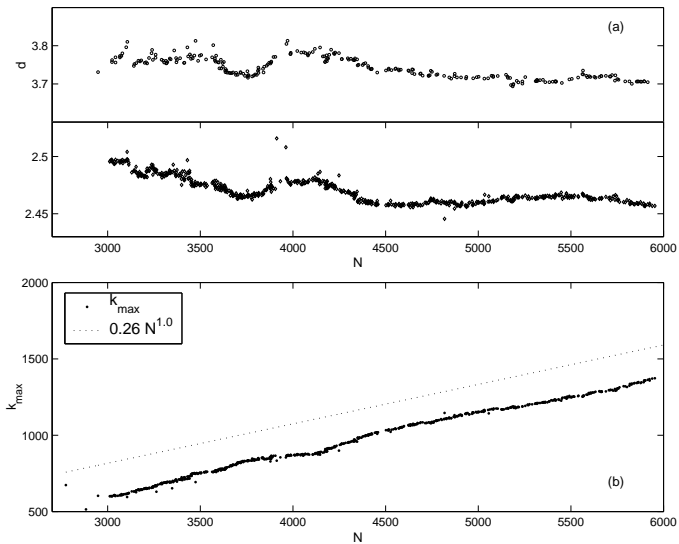}
\end{center}
\caption{(a) Evolution of the Internet backbone diameter with the
number of nodes recorded on NLANR Internet maps. The top curve is
the measured average shortest path on the graphs, while the bottom
panel shows the quantity computed from the degree distributions
using Eq. (\ref{newman}). Days where the mapping was incomplete
are omitted. (b) The maximum degree of a node scales linearly with
the size of the network. \label{moat}}
\end{figure}

These results are valid for the model of random graphs constructed
according to the simple method assumed in Ref.
\cite{newman01graphs}. Even for these ideal graphs, Eq.
(\ref{newman}) numerically underestimates the actual path lengths,
as is obvious from its method of derivation, but it does capture
the proper scaling behavior. Real-world graphs with additional
non-random structure can have even longer average shortest paths.
Nevertheless, as shown in
\cite{albert02review,newmanchapter,newman_scientific_collabII},
the formula does approximate the diameter for many real-world
graphs, especially with respect to the dependence of the average
path length on $N$. In our experiments on two diverse real-world
data sets, we also find the numerical underestimation. Still, a
comparison between the behavior of Eq. (\ref{newman}) and measured
diameters on our data sets for all $N$ shows that this analytical
treatment tracks the basic features of the data quite well (see
Figs. \ref{metabolic}a and \ref{moat}a).

We next consider our assumptions and results in the context of two
data sets, each representing snapshots of the topologies of graphs
that grow in an environment with possible selection pressures. The
first concerns the organization of essential biological processes
within the cell for a variety of simple organisms, and the second
concerns the large-scale structure of the Internet backbone. While
these are, of course, very different examples, we will show that
they do share features consistent with the arguments just
presented, and we provide reasons for why a selection argument
related to the diameter of these networks is reasonable.

Metabolic pathways are complex biological networks in which a
series of enzymatic reactions produce specific products within
cells. Large scale sequencing projects have furnished integrated
pathway-genome databases\cite{karp,kanehisa,wit} from which
organism-specific metabolic networks can be inferred. In a
metabolic network, nodes represent the substrates, and a directed
link connects the educt to the product of a metabolic reaction.

Recently, such databases have been used \cite {barabasi,fell} to
analyze the topological properties of the metabolic networks of
$43$ different organisms including \textit{E-coli} (bacterium) and
\textit{Caenorhibditis elegans} (eukaryote).  The network degree
distributions were found to be uniformly scale-free with exponents
between $2.0$ and $2.4$. A striking feature of the metabolic
networks studied is that even though their sizes vary between 200
and 800 nodes, the diameter stays approximately fixed between
$3.0$ and $3.5$, as shown in Fig. \ref{metabolic}a.

It has been speculated \cite{fell,fellproceedings} that metabolic
networks may have evolved to maintain a constant diameter in order
to minimize the number of sequential reactions necessary to obtain
a particular product. For example, it was found that there are
several possible pathways which could provide the same chemical
solution as the Krebs cycle, but the true Krebs cycle is the most
efficient and contains the least number of steps \cite{enrique}.
Another possible evolutionary force is opportunism, where a new
metabolic pathway is developed by re-using enzyme catalyzing
reactions already in the cell rather than developing an entirely
new pathway from scratch \cite{fell}. Thus, as the network
evolves, existing substrates are incorporated into new pathways
and their connectivity grows. Fig. \ref{metabolic}b shows that the
degree of the most connected node grows linearly with the size of
the metabolic network, in agreement with the assumption $k_c \sim
N^{\alpha}$.

Further potential benefits of small diameters include the
reduction in the transition time between metabolic states
\cite{fell} in response to environmental changes. Networks with
robustly small average path lengths have been found to rapidly
adjust to perturbations \cite{wattssw}. Thus there may be a
selective advantage to maintaining a small diameter.

\begin{figure}
\begin{center}
\includegraphics[scale=0.45]{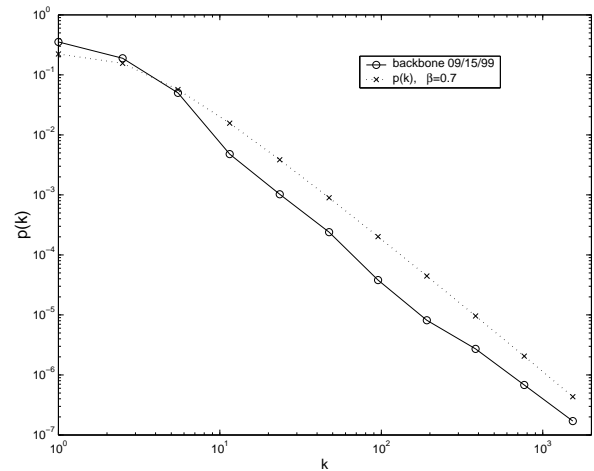}
\end{center}
\caption{An example degree distribution for the Internet backbone
compared with a distribution computed using the recursion in Eq.
(\ref{recurrence}). \label{backboneVStheoretical}}
\end{figure}

The Internet backbone is a second example of a network where
maintaining a constant diameter is important. Data is routed on
the Internet between tens of millions of host computers by
breaking the data up into packets, each of which are routed
individually and then re-assembled upon arrival at their
destination.  The packets hop from node to node in the network.
Each additional hop a packet must make introduces latency and
increases the potential for signal degradation through errors and
delays. We measured the diameter of Internet maps from November
1997 to January 2000 gathered by the National Laboratory for
Applied Network Research (NLANR)(\textit{http://moat.lanr.net}).
Each node is an autonomous system (AS) usually corresponding to a
single Internet Service Provider, and the links represent
inter-ISP connections. The Internet backbone connectivity
distribution is power-law with an exponent $\gamma = 2.2 \pm 0.1$
invariant over time
\cite{vespignaniPRLbackbone,faloutsos99topology}. An example
degree distribution is shown in Fig. \ref{backboneVStheoretical}
along with a distribution calculated using the recursion in Eq.
(\ref{recurrence}).  Fig. \ref{moat}a shows, consistent with
previous measurements \cite{vespignaniPRLbackbone}, that the
diameter stays approximately constant at 3.7 hops over the 2 year
period, while the number of nodes doubles from three to six
thousand. It appears that the Internet backbone may have evolved
to connect a greater number of ISPs but has kept the average
number of hops Internet traffic must make low.

In summary, we have presented a plausible reason for the existence
of scale-free distributions observed in two contexts, metabolic
networks and the Internet backbone, where there are evolutionary
pressures to maintain a small diameter. Our analysis shows that
for a robust network to maintain its diameter, the form of its
degree distribution should be scale-free. We have further shown
our assumptions to be consistent with observed features in the two
data sets. Combined with endogenous models of preferential
attachment, and the error tolerance of scale-free networks, our
results help further explain the prevalence of scale-free networks
in selective environments.

We acknowledge B. A. Huberman for many useful discussions and A.
R. Puniyani for contributions to earlier versions of this work.

\bibliography{lukoseavshortpath}
\end{document}